# Exploring Phononic Properties of Two-Dimensional Materials using Machine Learning Interatomic Potentials

Bohayra Mortazavi*[a,b], Ivan S. Novikov[c,d], Evgeny V. Podryabinkin[c], Stephan Roche[e,f], Timon Rabczuk[g], Alexander V. Shapeev**[c] and Xiaoying Zhuang[a,g]

[a]*Chair of Computational Science and Simulation Technology, Department of Mathematics and Physics, Leibniz Universität Hannover, Appelstraße 11,30157 Hannover, Germany.*
[b]*Cluster of Excellence PhoenixD (Photonics, Optics, and Engineering–Innovation Across Disciplines), Gottfried Wilhelm Leibniz Universität Hannover, Hannover, Germany.*
[c]*Skolkovo Institute of Science and Technology, Skolkovo Innovation Center, Nobel St. 3, Moscow 143026, Russia.*
[d]*Institute of Materials Science, University of Stuttgart, Pfaffenwaldring 55, 70569 Stuttgart, Germany*
[e]*Catalan Institute of Nanoscience and Nanotechnology (ICN2), CSIC and BIST, Campus UAB, Bellaterra, 08193 Barcelona, Spain.*
[f]*ICREA Institució Catalana de Recerca i Estudis Avancats, 08010 Barcelona, Spain*
[g]*College of Civil Engineering, Department of Geotechnical Engineering, Tongji University, Shanghai, China.*

## Abstract

Phononic properties are commonly studied by calculating force constants using the density functional theory (DFT) simulations. Although DFT simulations offer accurate estimations of phonon dispersion relations or thermal properties, but for low-symmetry and nanoporous structures the computational cost quickly becomes very demanding. Moreover, the computational setups may yield nonphysical imaginary frequencies in the phonon dispersion curves, impeding the assessment of phononic properties and the dynamical stability of the considered system. Here, we compute phonon dispersion relations and examine the dynamical stability of a large ensemble of novel materials and compositions. We propose a fast and convenient alternative to DFT simulations which derived from machine-learning interatomic potentials passively trained over computationally efficient ab-initio molecular dynamics trajectories. Our results for diverse two-dimensional (2D) nanomaterials confirm that the proposed computational strategy can reproduce fundamental thermal properties in close agreement with those obtained via the DFT approach. The presented method offers a stable, efficient, and convenient solution for the examination of dynamical stability and exploring the phononic properties of low-symmetry and porous 2D materials.
Keywords: Machine-learning; Interatomic potentials; Phononic properties; 2D materials;
Corresponding authors: *bohayra.mortazavi@gmail.com; **a.shapeev@skoltech.ru

## 1. Introduction

Phonon dispersion relations (PDRs) are key components to study the lattice dynamics and atomic vibrations in a crystal[1,2]. They also provide useful information regarding the transport properties such as the thermal conductivity. In particular, PDRs are extensively employed to examine the dynamical stability of various compositions and structures. To acquire PDRs, the most popular theoretical approach is to calculate the force constants using the density functional theory (DFT) simulations. To assess the thermal properties, DFT calculations are commonly executed using supercell structures. DFT is a computationally efficient approach for the majority of highly symmetrical lattices. In recent years, two-dimensional (2D) materials[3,4] are gaining remarkable attentions because of their unique properties, suitable to address critical challenges in various advanced technologies like nanoelectronics, energy harvesting devices or biosensors. Since the isolation of graphene[1,2], 2D materials family has been continuously and quickly extending. This large family of new materials includes highly symmetric lattices, such as the graphene-like structures [5,6] and 2H transition metal dichalcogenides[7,8], as well as many structures with low-symmetry or nanoporous lattices like graphdiyne[9] and 1T' transition metal dichalcogenides[10]. To date, high-symmetry and isotropic lattices have been excessively studied. Recently, however, anisotropic and nanoporous 2D lattices are attracting considerable attention, because they offer new possibilities to achieve angle-dependent properties and design more efficient energy storage/conversion systems.

In order to theoretically predict novel 2D systems, it is thus essential to carefully examine the dynamical stability and explore the vibrational properties on the basis of PDRs. For low-symmetry and nanoporous 2D structures, currently employed DFT simulations exhibit limited flexibility due to severe computational issues. In these cases, to bypass the computational limitations, one usually treats smaller supercells and lower k-point girds and plane-wave cutoff energy within the DFT simulations. But such simplifications may actually result in poorer accuracy of the acquired PDRs. Moreover, in some cases nonphysical imaginary frequencies may appear in the PDRs thus impeding the examination of dynamical stability or thermal properties. Machine-learning based methods are currently among the most attractive solutions to address various challenges, also in materials science[11–17]. In this regard, machine-learning interatomic potentials (MLIP) [18] have shown outstanding efficiency in many applications of computational material science, such as predicting novel materials [19,20] lattice dynamics[21] and estimating the thermal conductivity[22,23], etc. The main advantage of MLIPs is that they enable the efficient use of classical molecular dynamics simulations to evaluate the forces and energies with the DFT level of accuracy. The employment of MLIPs is particularly promising to study the large systems, for those DFT-based methods become infeasible due to the known computational limitations.

In this work we propose a computationally efficient and accurate methodology to acquire PDRs and explore other critical phononic properties on the basis of passively fitted moment tensor potentials (MTPs)[24]. MTP belongs to the family of MLIPs [25–27] by which we understand potentials possessing flexible functional form that allows for systematically increasing of the accuracy with an increase in the number of parameters and the size of the training[28]. Similar to other MLIPs, MTP is able to approximate any model of interatomic interaction, offering a compromisable choice between accuracy and computational efficiency. The proposed approach only requires inexpensive and short ab-initio molecular dynamics trajectories, without the need in active learning or additional DFT calculations. We employed such an approach to evaluate the phononic properties of a wide variety of 2D materials, for which close agreements with density functional perturbation theory (DFPT) results are observed. We describe our method on the basis of passively-trained MTPs, which can be conveniently employed to explore the phononic properties of low-symmetry and nanoporous 2D materials, with enhanced stability and computational efficiency.

## 2. Computational methods

First-principles DFT calculations in this work were carried out using the *Vienna Ab-initio Simulation Package* (VASP)[29–31]. The generalized gradient approximation (GGA), namely the Perdew–Burke–Ernzerhof (PBE)[32] functional was employed in the calculations. We assumed the plane-wave cutoff energies of 600 eV and 400 eV for the carbon-based and carbon-free systems, respectively. For geometry optimization, the convergence tolerances for the energy and forces were set to $10^{-5}$ eV and 0.001 eV/Å, respectively. The DFT simulations were performed over supercell samples using a 3×3×1 Monkhorst-Pack[33] k-point grid. The plane-wave cutoff energy of the DFPT simulations was set as the default value by VASP. The PHONOPY code[34] was utilized to create the optimal sets of atomic position for DFPT calculations and also to acquire phonon dispersions and group velocities with the DFPT results as inputs. Ab-initio molecular dynamics (AIMD) simulations were performed with a time step of 1 fs using a 3×3×1 k-point gird.

A class of machine-learning interatomic potentials—moment tensor potentials[24] were used to describe interatomic interactions. Similar to classical potentials, MTPs include parameters that are optimized on a set of training configurations. In this work AIMD simulations were used to create the training sets. MTP was first proposed for single-component systems[24] and has been recently generalized to multiple component systems[19,35]. This potential is local, i.e., the total energy $E$ of the system containing $N$ atoms is partitioned into contributions $V$ of neighborhoods $u_i$ of each $i$-t atom: $E \equiv E^{MTP} = \sum_{i=1}^{N} V(u_i)$. We refer to the $j$-th atom as a neighbor of the $i$-th (central) atom if the distance between them is less than a predefined cut-off distance $R_{\text{cut}}$. The neighborhood is then expressed as a collection,

$$u_i = \left(\{r_{i1}, z_i, z_1\} \dots, \{r_{ij}, z_i, z_j\} \dots, \{r_{iN_{\text{neigh}}}, z_i, z_{N_{\text{nei}}}\}\right), \quad (1)$$

where $r_{ij}$ are the relative atomic positions (interatomic vector), $z_i$ and $z_j$ are the types of the central and neighboring atoms, respectively, and $N_{\text{nei}}$ is the number of atoms in neighborhood. Each contribution to the total energy has the following form: $V(u_i) = \sum_\alpha \xi_\alpha B_\alpha(u_i)$, where $\xi_\alpha$ are the free parameters of the potential to be optimized, and $B_\alpha$ are the basis functions. We construct the basis functions as all possible contractions of the moment tensor descriptors:

$$M_{\mu,v}(r_i) = \sum_{j=1}^{N_{\text{nei}}} f_\mu(|r_{ij}|, z_i, z_j) r_{ij}^{\otimes v} \quad (2)$$

yielding a scalar (see Ref.[24] for details). The first factor $f_\mu(|r_{ij}|, z_i, z_j)$ in the aforementioned summation is the radial part depending only on the distance between atoms $i$ and $j$ and their types. We expand the radial part through a set of radial basis functions $\varphi_\beta(|r_{ij}|)$ multiplied by a factor $(R_{\text{cut}-}|r_{ij}|)^2$ for smoothing near the cut-off radius.

$$f_\mu(|r_{ij}|, z_i, z_j) = c_{\mu,z_i,z_j}^{(\beta)} \varphi_\beta(|r_{ij}|)(R_{cut-}|r_{ij}|)^2, \quad (3)$$

where $c_{\mu,z_i,z_j}^{(\beta)}$ are the radial coefficients. We denote by "$\otimes$" the outer product and refer to the second factor in Eq. 2 as the angular part. In order to optimize the $\xi_\alpha$ and $c_{\mu,z_i,z_j}^{(\beta)}$ parameters of an MTP, one needs to solve the following minimization problem (training of MTP):

$$\sum_{k=1}^{K}\left[w_e(E_k^{\text{AIMD}}-E_k^{\text{MTP}})^2 + w_f \sum_i^N |f_{k,i}^{\text{AIMD}}-f_{k,i}^{\text{MTP}}|^2 + w_s \sum_{i,j=1}^{3} |\sigma_{k,ij}^{\text{AIMD}}-\sigma_{k,ij}^{\text{MTP}}|^2\right] \to \min, \quad (4)$$

where $E_k^{\text{AIMD}}$, $f_{k,i}^{\text{AIMD}}$ and $\sigma_{k,ij}^{\text{AIMD}}$ are the energy, atomic forces and stresses in the training set, respectively, and $E_k^{\text{MTP}}$, $f_{k,i}^{\text{MTP}}$ and $\sigma_{k,ij}^{\text{MTP}}$ are the corresponding values calculated with the MTP, $K$ is the number of the configurations in the training set, and $w_e$, $w_f$ and $w_s$ are the non-negative weights that express the importance of energies and forces and stresses in the optimization problem, respectively, which in our study were set to 1, 0.1 and 0.001, respectively. We note that the weights for the energy and force are the default values.

Here we used the PHONOPY code[34] to evaluate the phononic properties, in which MTP replaces VASP in the force calculation step[36]. To facilitate future studies, in the data availability section we present a guide to create the training set (VASP AIMD inputs), as well as the MTP training procedure and the integration code to calculate the MTP forces of the PHONOPY input structures. In particular, all the examples of this work are included. We hope that such information will facilitate the use of MTP for accurate evaluation of the phononic properties in structural systems of versatile complexity.

## 3. Results and discussions

The main objective of this study is to replace DFPT with MTP in the force-constant calculations for the evaluation of phononic properties. Although our work is focused on the 2D materials, the proposed approach is equally valid for 3D or 1D lattices. First, we study the monoelemental 2D structures. Since carbon shows the unique ability to form diverse stable 2D atomic lattices, we mostly consider carbon-based structures. In Fig. 1, the studied monoelemental 2D lattices are illustrated. Apart from the graphene (Gr), we consider penta-graphene[37], various haeckelite[38] (Haeck) lattices, phagraphene [39] (Pha-Gr) and different graphyne [40] (GY) structures. We note that the haeckelite and phagraphene lattices include non-hexagonal carbon rings and can thus be considered as defective graphene lattices but with high degree of periodicities and close densities to that of the pristine graphene. On the other side, graphyne lattices are highly porous full carbon materials, which recently gained remarkable attentions because of experimental advances and their bright application prospect for the energy storage/conversion systems. Besides full-carbon structures, we also consider the single-layer black phosphorene (P) which has a rather complicated buckled lattice.

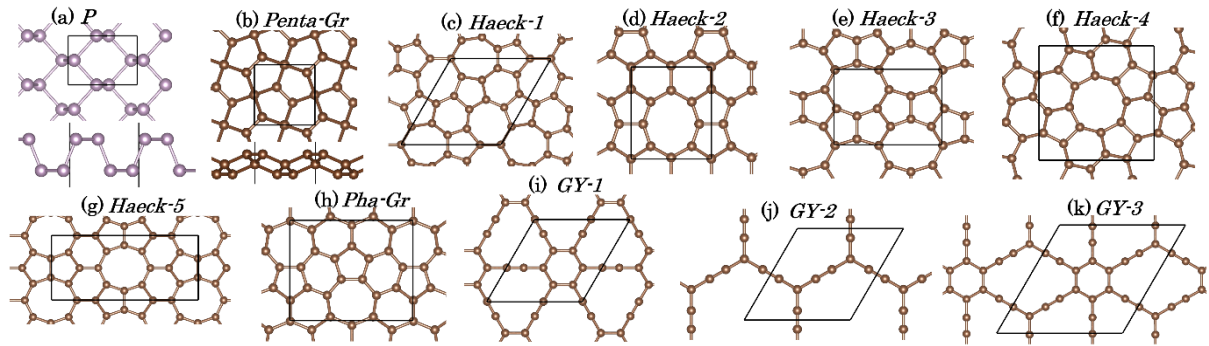

**Fig. 1**, Top views of atomic structures of the considered monoelemental systems. Side views are shown only for non-planar lattices. Black lines illustrate the primitive unit cell.

When using the DFPT method, the phononic properties are acquired on the basis of applying small displacements in supercell lattices. Therefore, the preliminary training data sets were prepared by conducting the AIMD simulations at a temperature of 50 K, with an overall simulation time of less than 1 ps (1000 simulation time steps). In this case, the AIMD trajectories include configurations with small lattice deformations from the equilibrium condition, consistent with our employed DFPT procedure. Since the MTPs are trained with a cutoff distance of 4 or 5 Å, the AIMD simulations were conducted over supercells with periodic box sizes over 10 Å. Larger supercells can be useful to describe the local large deflections better. Nonetheless, supercells with the minimum sizes closer to 10 Å are more computationally efficient because the costs of AIMD simulations increase exponentially with the number of atoms. To facilitate future studies, all created training sets in this work are provided in the data availability section. Using the AIMD results, we trained MTPs with 901 parameters for the monoelemental systems. The cutoff distance for graphene, other carbon allotropes and phosphorene were assumed to be 3, 4 and 5 Å, respectively. In Fig. 2 we compare the PDRs predicted by MTPs with those by the DFPT. Remarkably close agreement between the MTP and DFT results are obtained. This is a highly promising outcome since the developed MTPs are first attempts without any long sequence of optimization. Noticeably for the penta-graphene, haeckelite lattices and phosphorene, acoustic and optical branches are reproduced with a high level of accuracy. The same observation is also valid for graphyne lattices. Notably, in some cases (Haeck-5, P, GY-2 and GY-3) slight imaginary frequencies appear in the PDRs calculated with DFPT, which are not observed in the MTP-based results. These imaginary branches in the DFPT results originate from computational artifacts, and are not representative of dynamical instabilities[41–43] and might be removed via more precise setups for the calculations, such as increasing the supercell size, increasing plane-wave cutoff energy or improving the resolution of k-point girds. These modifications can be affordably examined for highly symmetrical lattices like graphene, but for less symmetrical and nanoporous lattices may substantially increase the computational costs and lead to complexity of the problem.

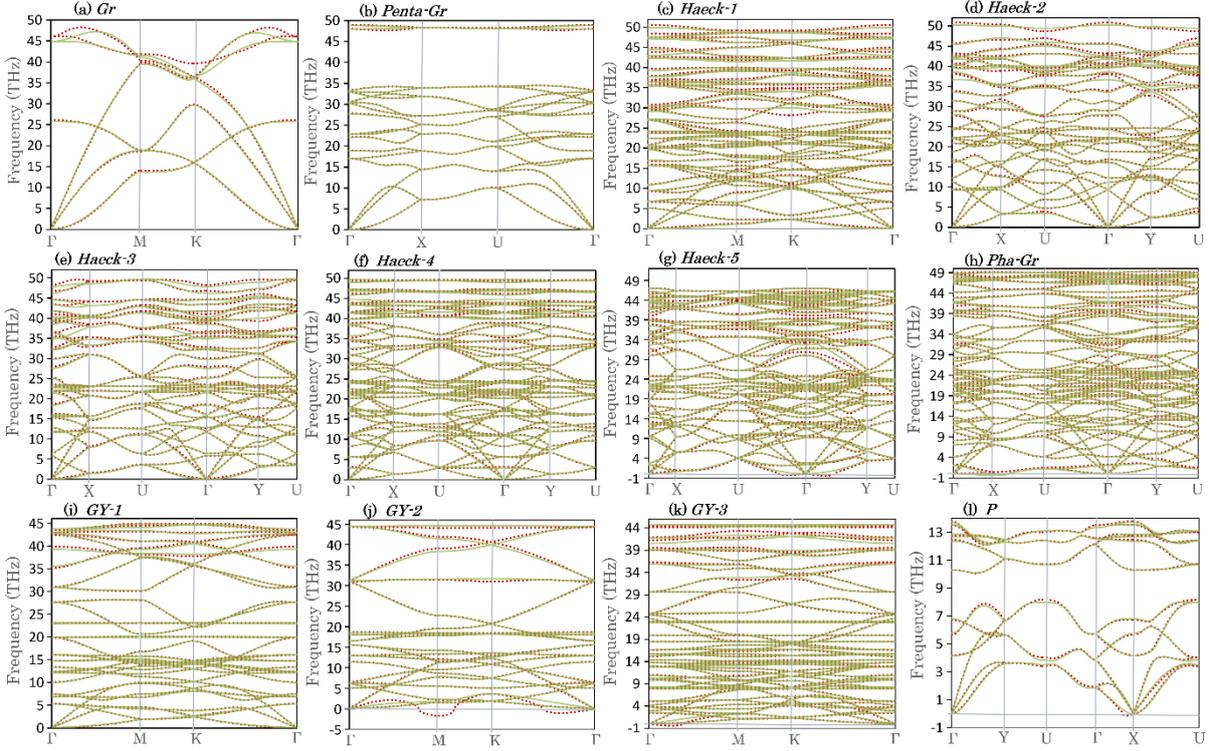

**Fig. 2**, Phonon dispersion relations of monoelemental 2D lattices acquired by the DFT (red-dotted lines) method and first-attempt MTPs (continuous green lines). The atomic lattices are show in Fig. 1.

Despite close agreement between the predicted PDRs for the most of monoelemental structures, two exceptions exist in which the MTP-based results show some inaccuracies. Among all the studied samples, it is clear that the high-frequency optical modes are well reproduced by the MTP. Unexpectedly for the highest symmetrical lattice of graphene, optical modes with the highest frequencies are not well reproduced by the MTP (Fig. 2a). On the other side for the phagraphene (Fig. 2h), a slight soft mode appears in the MTP result which is absent in the DFPT result. These disagreements between the DFPT and MTP results reveal that some degree of extrapolation may have occurred and accordingly suggest that for these samples the created passive training sets may not have been as accurate as those for other samples. We remind that when conducting the AIMD simulations at the low temperature of 50 K for a very short simulation time, the occurring configurations remain highly correlated and some critical configurations may never get explored in the training set. To address this issue, a simple effective solution is to expand the training set by including additional uncorrelated AIMD trajectories at higher temperatures. To that end, additional 1 ps AIMD trajectories were included by conducting the simulations at 200, 300, 500, 700 and 900 K (each one for 0.2 ps). This way the total AIMD trajectories include 2 ps, and the new training sets were created accordingly. In Fig. 3, we include the predicted PDRs for single-layer graphene and phagraphene by the improved training sets and compare the new results with the original results. It is clear that by incorporating higher temperature trajectories, the imaginary frequencies in the phagraphene's PDR vanishes, while the close agreement with DFPT results at higher frequencies is kept intact. Similarly, for the case of graphene, the accuracy of high-frequency optical modes is considerably improved without affecting the lower frequency optical and acoustic modes. The acquired results reveal that the incorporation of high-temperature trajectories can further enhance the accuracy and stability of MTPs. It is worth noting that in prior works, MLIPs have been successfully employed to accurately reproduce the phononic properties of graphene [44,45]. In comparison with these earlier works, the preparation of training sets in our approach is however more computationally efficient. In addition, when comparing with prior studies [44,45], our method could more accurately reproduce the acoustic modes in graphene's PDR, which is an important finding because these modes are the dominant heat carriers in graphene.

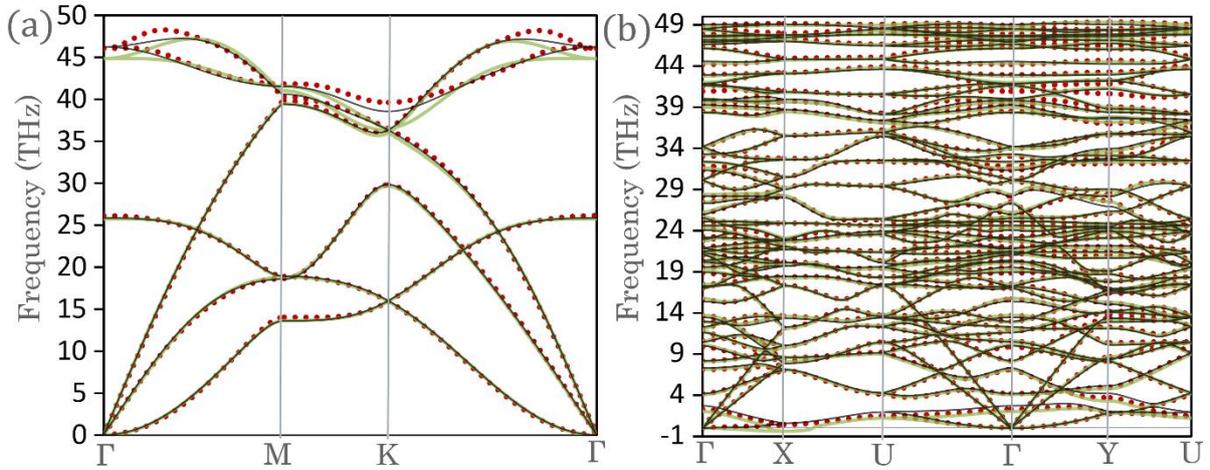

Fig. 3, Phonon dispersion relations of (a) graphene and (b) phagraphene acquired by the DFPT (red-dotted lines) and MTPs passively trained over AIMD trajectories at 50 K (continuous green lines) and 50 to 900 K (continuous green lines).

We next consider the binary 2D systems shown in Fig. 4. We note that various carbon-nitride 2D systems, like $CN$[41], $C_2N$[46], $C_3N_4$[47] and $C_3N$[48] are among the most attractive 2D semiconductors that have been experimentally fabricated and exhibit promising performances for a wide range of applications. $BC_3$[49] has been also experimentally fabricated, $C_4N$ and $C_7N_6$, $C_9N_4$ and $C_{10}N_3$ have been theoretically predicted by Li et al.[50] and Mortazavi et al.[51], respectively. We also consider carbon-free binary structures of hexagonal boron-nitride (BN), 2H transition metal dichalcogenides ($MX_2$, M= Mo, W and X=S, Se, Te), $SiP_2$[52] and $As_2Se_3$[53] monolayers. Similar to the case of monoelemental 2D lattices, we took MTPs with 1009 parameters and trained them on 1ps-long AIMD trajectories at the temperature of 50 K.

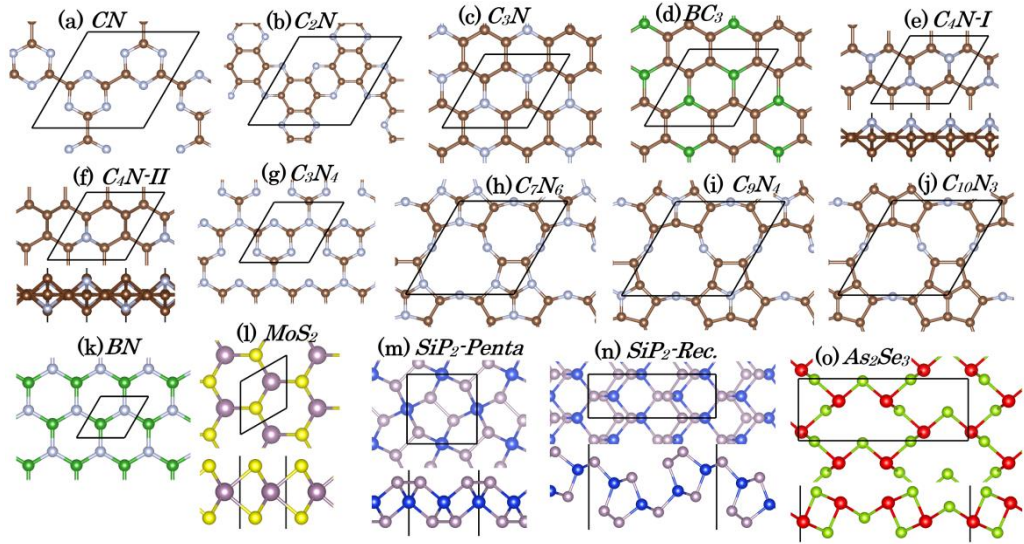

Fig. 4, Top views of atomic structures of binary 2D systems. Side views are shown only for non-planar lattices. Black lines illustrate the primitive unit cell.

The predicted PDRs by MTPs are compared with those by the DFPT method for the binary 2D lattices in Fig. 5. As it is clear, passively-trained MTPs can very accurately reproduce the PDRs for the considered binary lattices. It is noticeable that PDRs on the basis of MTPs are free of the slightly imaginary frequencies that have occurred around the Γ points in the DFPT results for some samples. The $C_3N_4$ monolayer with a full flat structure shows conspicuous imaginary frequencies, accurately reproduced also by the MTP-based method. These results are highly promising taking into account the complexity of the considered structures, especially in view of rectangular $SiP_2$ and $As_2Se_3$ exhibiting intricate buckled lattices. It should also be noted that while the acoustic modes are very closely reproduced for all considered samples, slight deviations are observed for some of the optical modes in a few samples, particularly for the $C_2N$ and $BC_3$ monolayers. As discussed for the case of graphene, to better reproduce the optical modes, the training set should be improved by including the AIMD trajectories at higher temperatures.

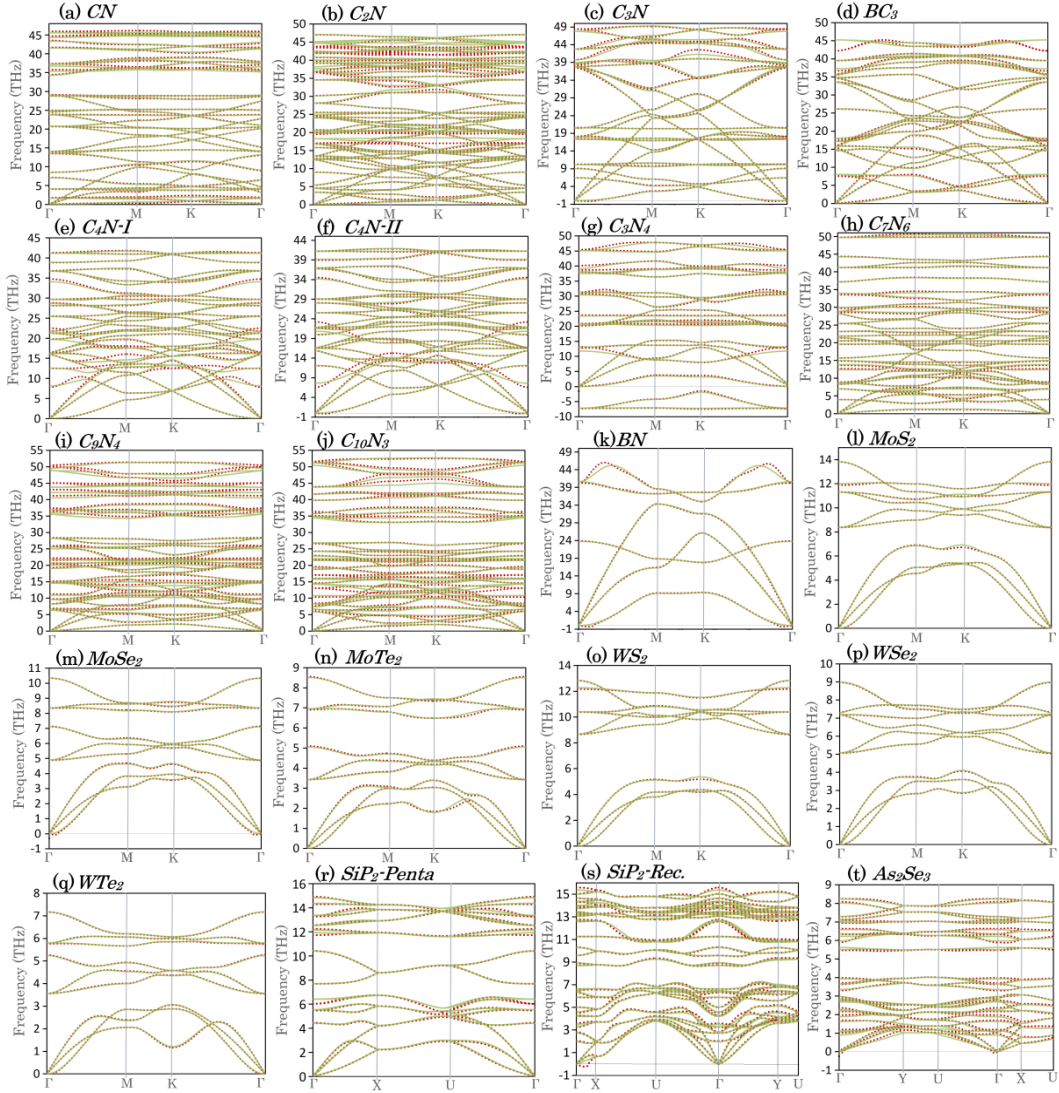

**Fig. 5**, Phonon dispersion relations of binary 2D lattices acquired by the DFPT (red-dotted lines) and first-attempt MTPs (continuous green lines). The corresponding atomic lattices are show in Fig. 4.

We next explore the accuracy of the PDRs predictions by the MTP method for ternary 2D lattices. In Fig. 6 we compare the PDRs for six different ternary lattices by the DFPT and MTP methods. We note that $BC_{10}N_2$ is recently predicted by Tromer et al.[54], $BC_6N$ shows hexagonal and rectangular atomic lattices, $BrCuTe_2$ and $ICuTe_2$ are also predicted according to their bulk counterparts, and $BC_6N_6$ is a novel ternary lattice whose stability we examine. It is noticeable that imaginary frequencies in the DFPT results are also well reproduced by the passively trained MTPs. These results, consistent with our previous observations, highlight that the accuracy of MTPs are insensitive to the number of elements' types and/or complexity of the structure. Although we have limited our AIMD simulations to less than 1 ps at 50 K, results presented up to this stage reveal that for the most of cases no additional AIMD trajectories are required to assess the phonon dispersions. Nonetheless, if required, the incorporation of additional high temperature AIMD trajectories in the training set is expected to improve the accuracy and stability as well.

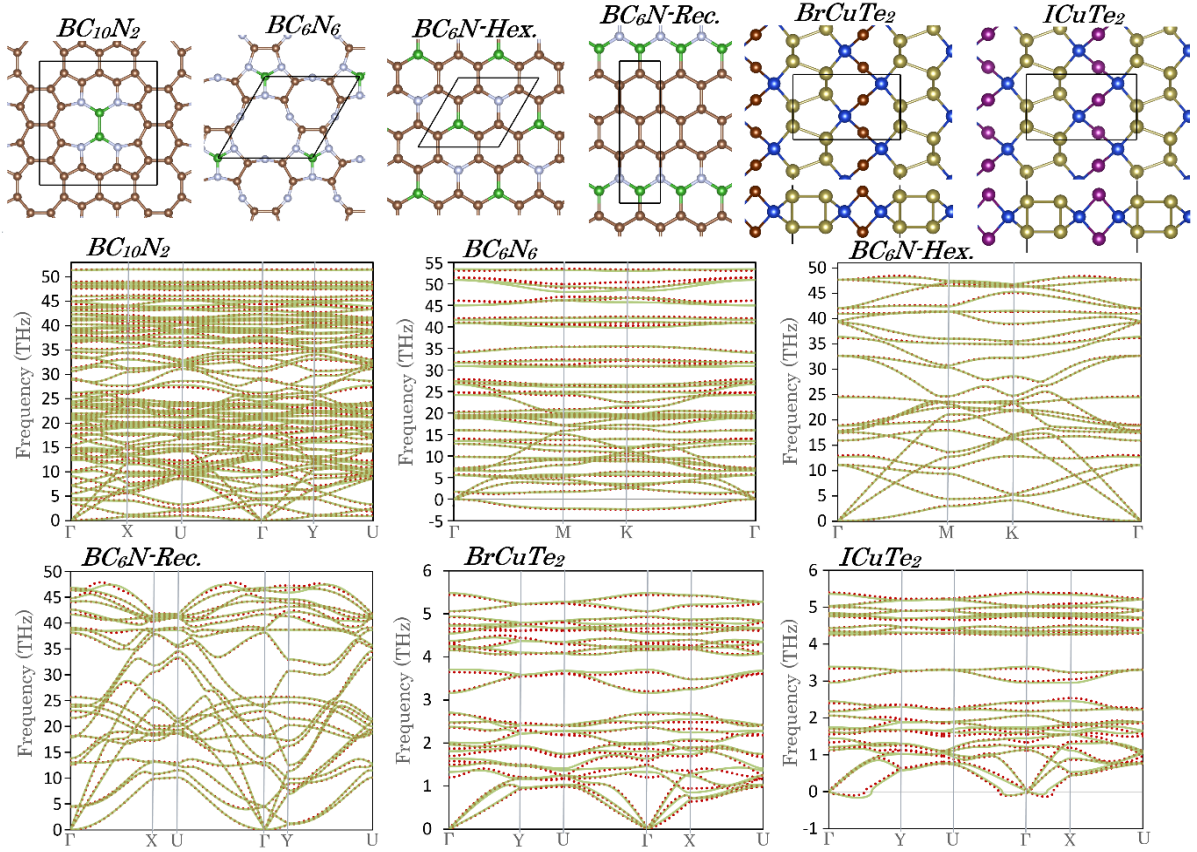

Fig. 6, Phonon dispersion relations for six different ternary 2D lattices acquired by the DFPT (red-dotted lines) and first-attempt MTPs (continuous green lines).

Besides the phonon dispersions, the phonons group velocity is another important phononic property, which can provide useful information concerning the lattice thermal conductivity. In Fig. 7 we compare the phonon group velocities predicted by the MTP method with those obtained within the DFPT using the PHONOPY code[34]. Remarkably, the passively trained MTPs can very closely reproduce the phonon group velocities for the studied samples. For some of the samples like graphene, penta-graphene and -SiP$_2$, CN, BC$_6$N$_6$ and transition metal dichalcogenides (MX$_2$, M= Mo, W and X=S, Se, Te) excellent agreement between the MTP method and DFPT-based estimations are observed. It is clear that in the worst cases, like BrCuTe$_2$, BC$_3$, or C$_2$N the group velocities are slightly shifted up or down and the results are free of substantial inaccuracies.

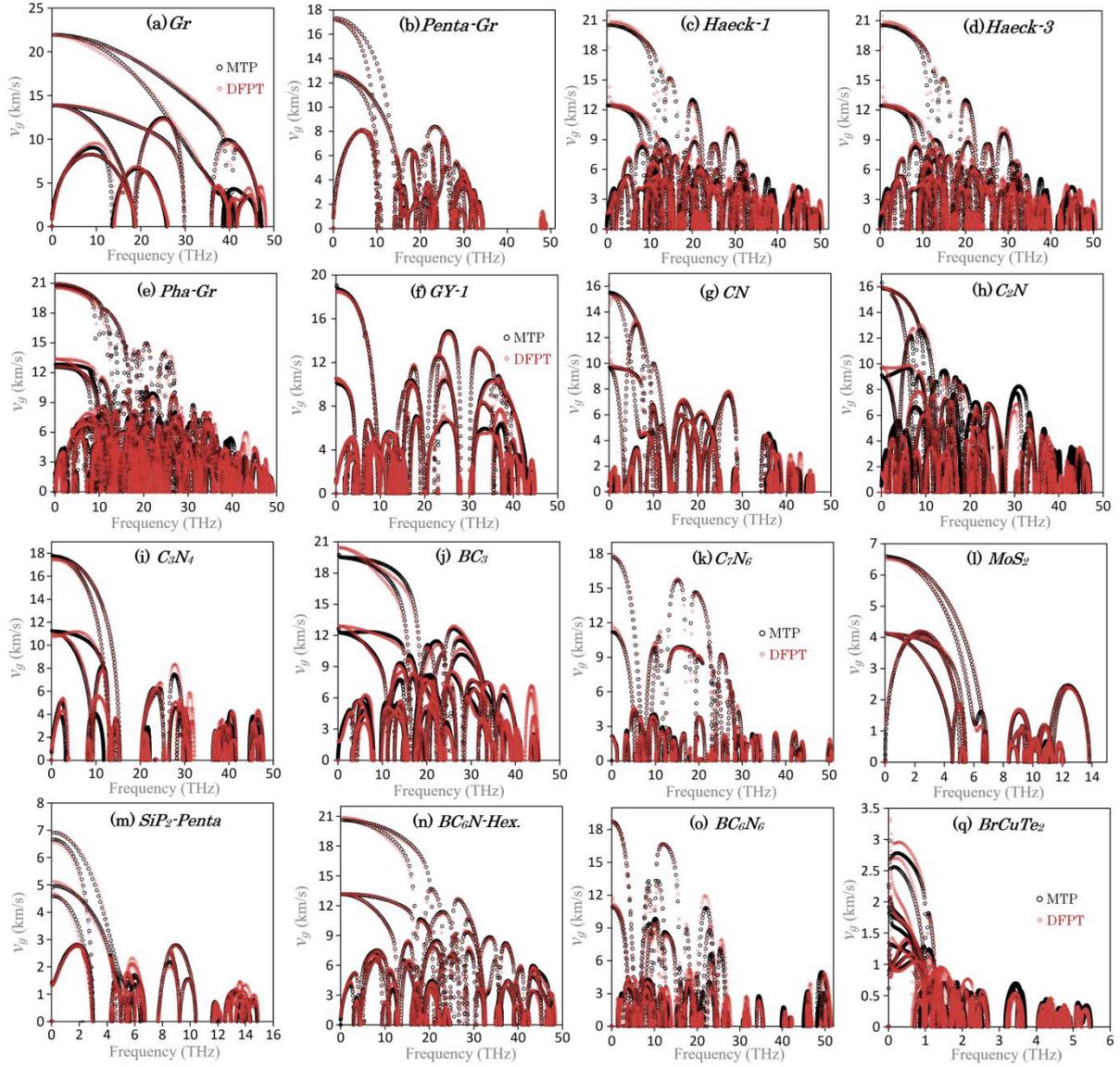

**Fig. 7**, Phonon group velocity ($v_g$) predicted by MTP and DFPT methods.

Last but not least, we examine the agreement between the MTP- and DFPT-based approaches to calculate the free energy, heat capacity, and entropy from their statistical thermodynamic expressions using the PHONOPY code[34]. We compared the MTP- and DFPT-based estimations for the aforementioned thermal properties in Fig. 8. As it is clear, for these results the agreement between DFPT and MTP results is excellent.

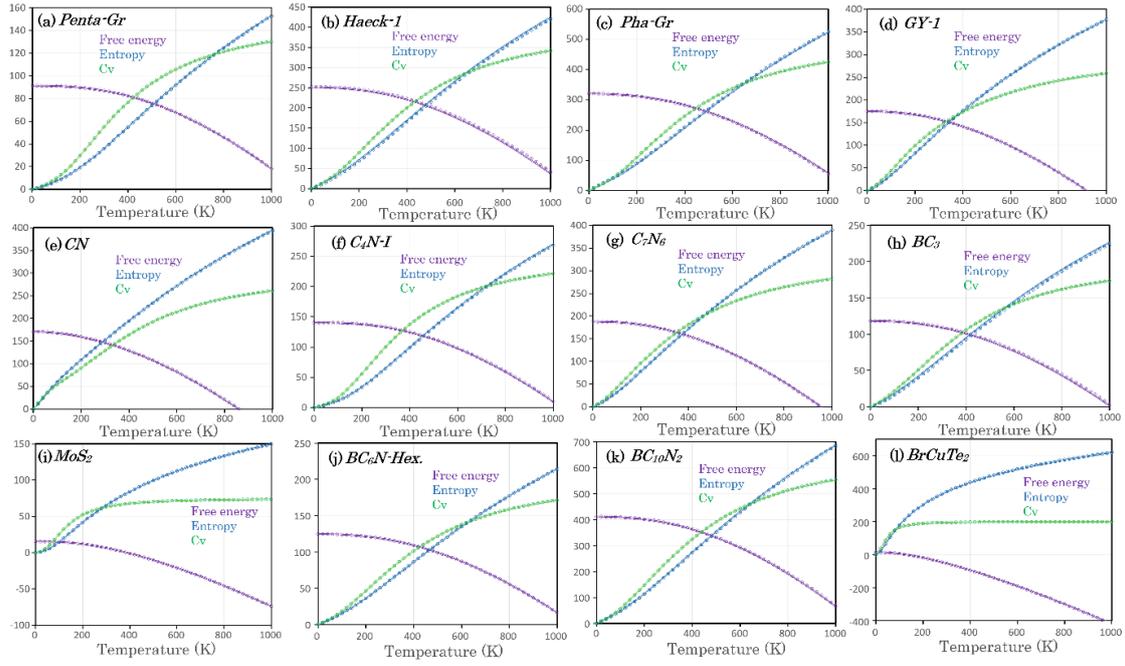

Fig. 8, Free energy, heat capacity, and entropy with the units of kJ/mol, J/K/mol, and J/K/mol, respectively, on the basis of MTP (continuous lines) and DFPT (circles) methods acquired using the PHONOPY code[34].

Presented results highlight the accuracy and stability of the MTP-based method to explore the phononic properties of complex 2D lattices. The required AIMD simulations to assess the phononic properties can be conducted relatively fast and are computationally less expensive than those commonly performed to examine the thermal stability (which usually requires more than 10 ps-long AIMD simulations). In comparison with DFPT method, AIMD simulations show lower sensitivity to the K-point grid and plane-wave cutoff energy. Moreover, the computational cost for the examination of supercell size effect on the PDRs is marginal for MTP-based approach, which is in contrast the bottleneck of DFPT method. One of the most salient advantages of the MTP-based method is that it typically yields smooth dispersions for the acoustic modes near the Γ-point. Differently when using the DFPT method for 2D materials, the convergence of flexural acoustic mode near the Γ-point may turn into a difficult task [55]. Nonphysical imaginary frequencies in the acquired PDRs may result in the instabilities for evaluating other thermal properties, particularly when calculating the lattice thermal conductivity. Worthy to point out that one of the reasons for appearance of imaginary frequencies in the PDRs might stem from slight inaccuracies in the atomic lattice. When using the DFPT method, small structural changes in the primitive unit cell, either in the atomic position or strain in the lattice size, demand entirely new calculations, whereas these effects can be checked with negligible computation costs using the MTP-based method. We should also observe that for highly symmetrical structures like graphene, transition metal dichalcogenides and penta-graphene, the standard DFPT method is by a large extent more computationally efficient. Actually the computational efficiency of MTP significantly enhances as the symmetry decreases, like that for haeckelites and $C_7N_6$ structures or for materials with large primitive unit cells, such as: metal- or conductive-organic frameworks and graphyne/graphdiyne lattices. In the previous works[44,45] it was demonstrated that MLIPs can accurately reproduce the phononic properties of graphene as compared with the standard DFPT method, but with substantially higher computational costs, questioning the practical application of MLIPs for this purpose. However, our results for extensive 2D systems clearly confirm that for low-symmetry and highly porous atomic lattices, the MTP-based approach offers an accurate, simple and convenient alternative to commonly employed DFPT approach to access the phononic properties. This is therefore a novel application area of MLIPs in the materials science as they can accurately and efficiently substitute the standard DFPT method for the valuation of phononic properties of complex structures. Although, the majority of our results on the basis of short AIMD trajectories at the low temperature of 50 K reveal outstanding accuracy, it is nonetheless highly recommended to include AIMD trajectories at higher temperatures in the training set to ensure improved stability and accuracy as well. Moreover, to more accurately reproduce the high frequency optical modes, employment of larger supercells in the AIMD calculations is highly recommended. In this case, the increase in the computational cost can be partially compensated by shortening the AIMD simulation. In this work PBE method was employed in the AIMD calculations for the evaluation of phononic properties. However we note that the proposed MTP-based approach can be also effectively used for other exchange-correlation functionals.

## 4. Conclusion

Our extensive results for a wide variety of 2D materials highlight that machine-learning interatomic potentials trained over short ab-initio molecular dynamics trajectories are able to reproduce the phononic properties in a strikingly close agreement with those by DFPT-based results. The proposed methodology could therefore play a pivotal role to conveniently explore the phononic properties of a large variety of low-symmetry and porous nanomembranes, with a high level of accuracy and reproducibility. In recent years astonishing advances have been achieved on the synthesis of low-symmetry and highly porous atomic lattices, like metal- or conductive-organic frameworks and graphdiyne nanomembranes. It is clear that phononic properties of these novel systems can now be effectively explored using the approach proposed in this work, which would otherwise require excessive computational resources with DFT-based methods. Our study therefore proposes that MTP-based approach can accurately and efficiently substitute the standard DFPT method for the valuation of phononic properties of complex structures. To facilitate the practical application, the full computational details of the proposed approach are included in the data availability section.

### Acknowledgment

B.M. and X.Z. appreciate the funding by the Deutsche Forschungsgemeinschaft (DFG, German Research Foundation) under Germany's Excellence Strategy within the Cluster of Excellence PhoenixD (EXC 2122, Project ID 390833453). E.V.P, I.S.N., and A.V.S. were supported by the Russian Science Foundation (Grant No 18-13-00479). ICN2 is supported by the Severo Ochoa program from Spanish MINECO (Grant No. SEV-2017-0706) and funded by the CERCA Programme/Generalitat de Catalunya.

## Data availability

The following data are available to download: (1) a brief guide for the installation of MLIP package, (2) all energy minimized lattices, (3) passive training approach and related commands, (4) VASP input script for the AIMD simulations, (5) VASP output file (vasprun.xml) for the DFT calculations, (6) created training sets and trained MTPs (p.mtp), (7) the C++ code to calculate the force constants using the PHONOPY and MTP for the input geometries and force calculator, respectively and (8) PHONOPY related scripts to adjust the outputs, from: GITLAB: https://gitlab.com/ivannovikov/mlip_phonopy and Mendeley dataset of: http://dx.doi.org/10.17632/7ppcf7cs27.1 .